\documentclass[12pt]{iopart}
\newcommand{\eg}{{\it e.g.\,}}
\newcommand{\ie}{{\it i.e.\,}}
\newcommand{\etc}{{\it etc.\,}}
\newcommand{\pc}{p_{\rm c}}
\newcommand{\f}{\frac}
\newcommand{\outness}{D}
\newcommand{\outnessw}{W}
\newcommand{\bea}{\begin{eqnarray}}
\newcommand{\eea}{\end{eqnarray}}
\usepackage{graphicx}
\usepackage{amssymb}

\begin{document}
\title[Directed network modules]
{Directed network modules}
\author{Gergely Palla$^1$, Ill\'es J. Farkas$^{1,2}$, P\'eter Pollner$^1$, 
Imre Der\'enyi$^2$ and Tam\'as Vicsek$^{1,2}$}
\address{$^1$ Statistical and Biological Physics Research Group of HAS and}
\address{$^2$ Dept. of Biological Physics, E\"otv\"os Univ.,
  1117 Budapest, P\'azm\'any P. stny. 1A}

\begin{abstract}
A search technique locating 
network modules,
\ie, internally densely connected groups of nodes
in {\it directed networks} is introduced by extending 
the Clique Percolation Method originally proposed for undirected networks.
After giving a suitable definition for directed modules 
 we investigate their percolation transition 
in the Erd\H os-R\'enyi graph both analytically and
 numerically. We also 
analyse four real-world directed networks, including Google's own 
web-pages,
an email network, a word association graph and 
the transcriptional regulatory network of the yeast Saccharomyces cerevisiae.
The obtained directed modules are validated 
by additional information 
available for the nodes. 
We find that directed modules of real-world graphs inherently overlap 
and the investigated networks can be classified into two major groups
 in terms of the overlaps between the modules. Accordingly, 
in the word-association network and
 Google's web pages  overlaps are likely to contain in-hubs,
whereas the modules in the email- and transcriptional regulatory network
 tend to overlap via out-hubs.
\end{abstract}
\maketitle

\section{Introduction}
\label{sec:intro}

A widespread approach 
to the analysis of complex natural, social and technological
phenomena is to assemble the participating molecules, 
individuals or electronic devices 
and their interactions
into a network (nodes and links) and to 
infer functional characteristics
of the entire system from
this static web of connections \cite{b-a-revmod,dorog-mendes-book}.
This approach is rooted in, among others, 
statistical physics, where often 
the thermodynamic limit ($N\rightarrow\infty$, where $N$ is the number of 
nodes) is considered,
and the overall (large-scale) structure of connections is studied rather 
than the details at the level of nodes and links.
Accordingly, over the past few years, several broadly studied
{\it large-scale} properties of real-world webs 
have been uncovered, \eg, a low average distance 
combined with a high average clustering coefficient \cite{WS},
the broad (scale-free) distribution of node degree 
(number of connections of a node)
\cite{F3,BA99,latora06review,jeong00metabolic}
and various signatures of hierarchical/modular organisation
\cite{Ravasz02,Han04}.
In addition, detailed analyses of the
{\it small-scale} behaviour of the same complex webs
have revealed overrepresented local structures:
graph motifs \cite{milo02,milo04},
\ie, small groups of nodes (typically of size $3-5$) with
specifically arranged connections among them.
The identified small- and large-scale properties are both closely
related to the dynamical behaviour of the corresponding complex system.
Nodes with many connections (hubs)
often have a central role in traffic \cite{guimera_air}, 
while motifs act as building blocks
performing distinct basic information processing tasks \cite{alon_FFL}.

The {\it inter mediate-scale} substructures in networks (units
 larger than motifs), made up of vertices more densely
connected to each other than to the rest of the network, are often
referred to as communities, modules, clusters or cohesive groups
 \cite{scott-book,pnas-suppl,everitt-book,knudsen-book,gn-pnas,newman-europhys,rg,palla05nature} with no widely accepted, unique definition.
In the various types of networks these groups can represent, 
\eg, communities of people \cite{scott-book,watts-dodds,our_new_nature}, 
functional units in biology \cite{Ravasz02,sm} 
and set of tightly coupled stocks or industrial sectors in economy
\cite{onnela-taxonomy}. A reliable method to pinpoint network modules
has many potential industrial application, \eg , it can help service providers 
(phone, banking, web, \etc)
identify meaningful groups of customers (users), or support biomedical 
researchers in their search for individual target molecules and novel 
protein complex targets \cite{novel-complexes,novel-complex-function}. 
In addition, modules, and also some small subgraphs,
are appropriate for ``coarse-graining'' complex networks: 
each module/subgraph can be represented as a node and two such node can
be linked,
if the corresponding modules/subgraphs are connected (or overlap)
\cite{palla05nature,makse05nature,pollner}.

The key requirements towards network module search techniques
\cite{palla05nature,Kosub,newman06pnas} are that they should be local,
based on link density, and error-tolerant (the removal or insertion of a link
may alter only nearby modules). Furthermore, as
dense groups in real-world graphs often overlap with each other, 
the module finding methods should allow overlaps between the groups. For 
example, in a social web each person
belongs to several groups (family, colleagues and friends), 
in a protein interaction network
each protein participates in multiple complexes  \cite{CYGD} and a 
large portion of web-pages
is classified under multiple categories \cite{dmoz}.
Prohibiting overlaps during 
module identification strongly increases the percentage of 
false negative co-classified pairs.
As an example, in a social web
a group of  colleagues might end up in different modules, each
corresponding to their families, and, in this case, the network 
module corresponding 
to their work unit is bound to become lost. 

\begin{figure}[t!]
\centerline{\includegraphics[angle=0,width=0.65\columnwidth]{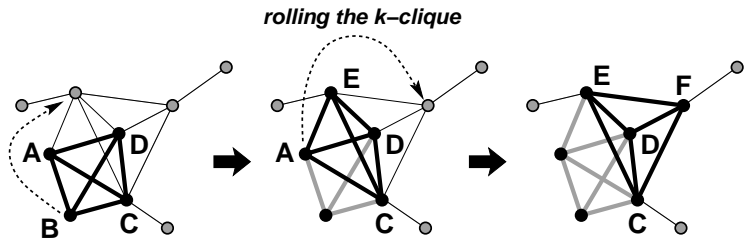}}
\caption[]{
Illustration of the Clique Percolation Method (CPM) 
\cite{palla05nature,DerenyiPRL} with $k$-clique template rolling 
in a small undirected graph for $k=4$. 
Initially the template is placed on A-B-C-D (left panel) and it 
is ``rolled'' onto the subgraph
A-C-D-E (middle panel). The position of the 
$k$-clique template is marked with thick black lines and black nodes,
whereas the already visited links are represented by thick gray lines
 and dark gray nodes.
Observe that in each step 
only one of the nodes is moved and the 
two $4$-cliques (before and after rolling) share $k-1=3$ nodes.
At the final step (right panel) the template reaches  the subgraph 
C-D-E-F, and the set of nodes visited during the process
(A-B-C-D-E-F) are considered as a module identified by the CPM at $k=4$.
}
\label{fig_CPM}
\end{figure}

A recent link-density based approach
 to module finding, fulfilling the above requirements,
is provided by 
the Clique Percolation Method (CPM)
\cite{palla05nature,DerenyiPRL}.  
In this approach, the definition of the modules is based on $k$-cliques 
(complete subgraphs of size $k$ in which each
 node is connected to every other node). A $k$-clique is a sub-graph 
with maximal possible link density, therefore
 it is a good starting point for defining modules. However, a method
accepting only complete sub-graphs as modules would be too restrictive.
Therefore, $k$-cliques are ``loosen up'' in the following way. 
Two $k$-cliques are said to be adjacent if they share $k-1$ nodes
 (or in other words, if they differ only in a single node), and a module
 is defined as the union of $k$-cliques that can be reached from each other
through a series of adjacent $k$-cliques. 
 Such modules can be best 
visualised with the help of a $k$-clique template (an object isomorphic
 to a complete graph of $k$ vertices). Such a template can be placed
onto any $k$-clique in the graph, and rolled to an adjacent
 $k$-clique by relocating one of its vertices and keeping its
 other $k-1$ vertices fixed. Thus, the $k$-clique modules 
($k$-clique communities) of a graph are all those
 subgraphs that can be fully  explored by rolling a $k$-clique template
in them, but cannot be left by this template, as illustrated 
in Fig.\ref{fig_CPM}. 
The algorithm used for the implementation of this technique is very
efficient for most real networks, and provides the full list of
overlapping modules in a short amount of time
\cite{palla05nature,our_bioinf}.

A common shortcoming of current module finding methods is that
they ignore the possible {\it directionality} of the links during
 the analysis of a network. The direction of a single link in most
 real network signals either the direction of some kind of flow (\eg ,
  the flow of information, energy), or the asymmetry of the 
relation between the nodes (\eg, a superior-inferior relation).
Consequently, nodes possessing mostly incoming links are expected
to play a very different role in the network (or within the modules they
belong to) from those possessing mostly outgoing links or from those
having a similar amount of both kinds of links. Therefore,
as a first attempt to take into consideration the directionality of
links, we propose a simple measure for the nodes within the modules to
characterise their roles in terms of the numbers of their incoming and
outgoing links.
%

At the same time the consideration of directionality in modules raises
the question of whether a module searching algorithm that inherently
takes into account the directionality of links is more suitable for
directed networks than the original undirected algorithms. Along this
idea, we define the notion of directed $k$-cliques (in which the
configuration of the directed links has to meet certain criteria), and 
propose a restricted version of CPM (denoted as CPMd), in which only
directed $k$-cliques can be used for the identification of modules.
We apply this method to several
networks: first, we
examine the percolation transition of the directed $k$-cliques in the 
Erd\H os-R\'enyi (ER) random graph \cite{ER}, then move on to
study  the directed modular structure of four real-world networks,
including a word-association network, Google's web-pages, an email network,
 and the transcriptional regulatory graph of yeast.
The identified directed modules are verified with the help of 
additional information (protein functional annotations, web-page names, and 
word usage frequencies) about the nodes.

\section{Definitions}
In undirected graphs a pair of nodes is either connected or not, 
whereas in a directed graph the same pair, (A,B),
can be connected in three ways: either by a ``single link'' as
(i) A$\rightarrow$B and (ii) A$\leftarrow$B or by a ``double link'' as
(iii) A$\rightleftharpoons$B.
Multiple links (\ie, more than one link between A and B in the same direction)
and self-links (such as A$\rightarrow$A) are not allowed. In the following we
 first define a simple measure for comparing nodes within a module based 
on the directionality of their links, then introduce the concept of directed
 $k$-cliques, the fundamental objects of our directed module finding approach.

\subsection{Comparing the nodes according to their relative out-degree}
A natural and simple approach to relate nodes in a module to each other is
 to compare the number of their incoming and outgoing links connected to other
 members in the module. For example, a node having only out-neighbours
 amongst the members of the module can be viewed a ``source'' or a ``top-node'',
 whereas a node with only incoming links from these members is a ``drain'' or
 a ``bottom-node''. Most nodes, however, fall somewhere between these
two extremes. To quantify this property, we introduce the 
{\it relative in-degree} and {\it relative out-degree} of node 
$i$ in module $\alpha$ as
\numparts
\bea
\outness_{i,{\rm in}}^{\alpha}&\equiv&\f{d_{i,{\rm in}}^{\alpha}}{d_{i,{\rm in}}^{\alpha}+
d_{i,{\rm out}}^{\alpha}},
\label{eq:innes} \\
\outness_{i,{\rm out}}^{\alpha}&\equiv&\f{d_{i,{\rm out}}^{\alpha}}{d_{i,{\rm in}}^{\alpha}+
d_{i,{\rm out}}^{\alpha}},
\label{eq:outness}
\eea
\endnumparts
where $d_{i,{\rm in}}^{\alpha}$ and $d_{i,{\rm out}}^{\alpha}$ denote the number of
 in-neighbours and out-neighbours amongst the other nodes in the module,
 respectively. Obviously the values of both $\outness_{i,{\rm out}}^{\alpha}$ and 
$\outness_{i,{\rm in}}^{\alpha}$ are in the range between 0 and 1, and the
 relation $\outness_{i,{\rm in}}^{\alpha}+\outness_{i,{\rm out}}^{\alpha}=1$ holds. 
For weighted networks,
 (\ref{eq:innes},\ref{eq:outness}) can be replaced by the {\it relative
   in-strength} and {\it relative out-strength} defined as 
\numparts
\bea
\outnessw_{i,{\rm in}}^{\alpha}&\equiv&\f{w_{i,{\rm in}}^{\alpha}}{w_{i,{\rm in}}^{\alpha}+
w_{i,{\rm out}}^{\alpha}},\label{eq:innesw}\\
\outnessw_{i,{\rm out}}^{\alpha}&\equiv&\f{w_{i,{\rm out}}^{\alpha}}{w_{i,{\rm in}}^{\alpha}+w_{i,{\rm out}}^{\alpha}},
\label{eq:outnessw}
\eea
\endnumparts
where $w_{i,{\rm out}}^{\alpha}$ and $w_{i,{\rm out}}^{\alpha}$ denote the 
aggregated weight of out-going and incoming  connections with 
other members in the module $\alpha$. 

\subsection{Directed $k$-cliques and the directed Clique Percolation Method (CPMd)}
\label{sect:directed_k_cliques}

\begin{figure}[t!]
\centerline{\includegraphics[angle=0,width=0.65\columnwidth]{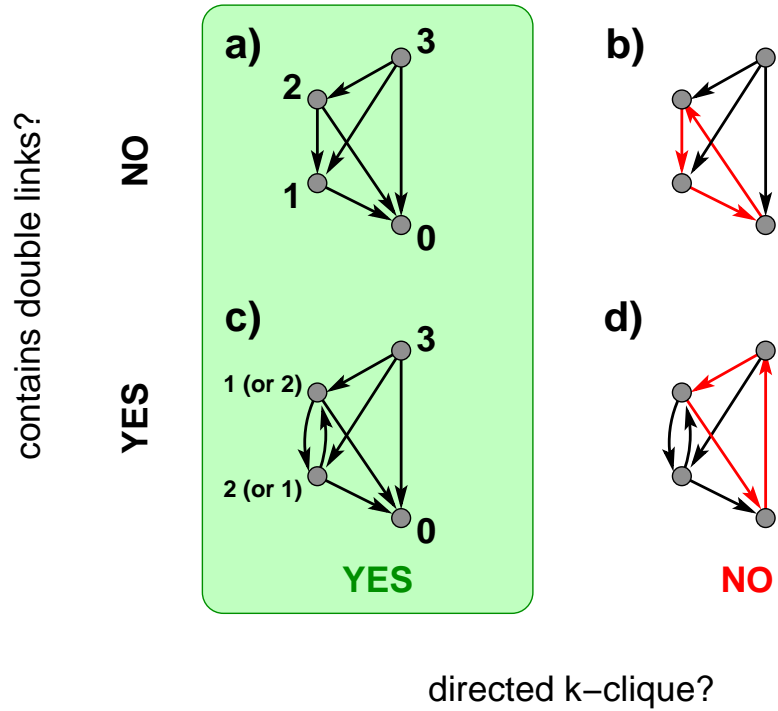}}
\caption[]{
Groups of nodes forming a directed $k$-clique
{\bf (a, c)}
and groups {\bf (b, d)} that do not.
{\bf (a)}
A directed $k$-clique without double links.
The index of each node corresponds to
its order (which is equivalent to number of its out-links) 
within the directed $k$-clique.
{\bf (b)} A complete sub-graph without double links,
but not accepted as a directed $k$-clique,
because it contains a directed loop.
{\bf (c)} A directed $k$-clique with a double link.
Note that the order of the nodes depends
on which link is deleted from the double link.
{\bf (d)} Double link in a complete sub-graph
that is not a directed $k$ clique.
It is not possible to remove a link from the double link
in a way that all directed loops disappear.
}
\label{fig:CPMd}
\end{figure}

In a complete sub-graph of size $k$ the $k(k-1)/2$ links can be 
directed in $3^{k(k-1)/2}$ ways.
Since the undirected CPM treats these  alternatives
as identical, introducing link directions allows a large variety of
possible rules for defining directed modules.
A natural concept, however, is to aim for ``directed modules''
 preserving some kind of directedness
 as a whole, rather than just being a collection of nodes 
 connected by directed edges. 

Therefore, we replace the $k$-cliques (the fundamental objects of the
CPM) by {\it directed $k$-cliques}, which are defined as complete sub-graphs of
 size $k$ in which an ordering can be made such that between any pair 
of nodes there is a directed
link pointing from the node with the higher order towards the lower one. 
Since the presence of double links usually leads to multiple possibilities to
 order the nodes in a way fulfilling the above requirement, for 
simplicity we first 
concentrate on directed $k$-cliques with no double links.
In this case, the higher the order of a node, the more out-neighbours it
has in the $k$-clique (see illustration in Fig.\ref{fig:CPMd}a). 
Thus, the {\it restricted out-degree} of a node in the
$k$-clique (the number of its out-neighbours in the $k$-clique, 
ranging from 0 to $k-1$) can be assigned as its order.
From this, it can be seen easily (for details see Appendix
A) that the condition for a $k$-clique with no double links to qualify
as a directed $k$-clique is equivalent to the following three
conditions:
\begin{itemize}
\item[(i)] Any directed link in the $k$-clique points from a node with a
  higher order (larger restricted out-degree) to a node with a 
lower order. 
\item[(ii)] The $k$-clique contains no directed loops
 (where a ``directed loop'' is a closed directed path).
\item[(iii)] The restricted out-degree of each node in the $k$-clique 
is different.
\end{itemize}
The overall directionality of such an object 
naturally follows the ordering of the nodes: 
the node with highest order is the one which has only 
out-neighbours, and can be viewed as the 
``source'' or ``top''-node of the $k$-clique, whereas the node with 
lowest order has only incoming
 links from the others, and corresponds to a ``drain'' or ``bottom'' node.

None of the above three conditions holds in the presence of double links:
 directed loops appear in the $k$-clique, the restricted out-degree of at 
least two nodes in the $k$-clique becomes the same (see Appendix A), and
we can find directed links pointing in the direction
 of increasing order. However, based on the ordering of the nodes,
it is always possible to eliminate the double links (by removing all
links that point towards higher order) from a directed $k$-clique in
such a way that the remaining single links fulfil all three conditions.
See Fig.\ref{fig:CPMd}c as an example.

The $k$-clique adjacency is defined similarly to the undirected case: 
two directed $k$-cliques are adjacent if they share $k-1$ nodes. The 
directed $k$-clique modules (the CPMd modules) arise as the union
of directed $k$-cliques that can be reached from each other through
 a series of $k$-clique adjacency. The $k$-clique template rolling picture can
 be applied to illustrate the CPMd modules in the same fashion as in the
 undirected case. The searching algorithm locating the CPMd modules is 
 described in Appendix B.


We note that the above definition of a directed $k$-clique is just one
possibility among many others. Natural choices that also impose some
kind of directionality on the $k$-clique include \eg the requirement
that at least one of the nodes should have out-links (or in-links)
towards (from) all the other $k-1$ nodes, or the requirement that the
nodes could be divided into two non-empty sets such that each node in
the first set has an out-link towards each node in the second set
(resembling directed hyper-edges). Our particular choice was motivated,
on the one hand, by the fact that it is more restrictive than the
others (providing a more specific tool to investigate the effects of
directionality) and, on the other hand, by our finding that for most
real world networks even such a restricted definition results in
directed modules that are notably similar to the undirected ones 
(see Sec.\ref{subsection:compare}).

\section{Percolation transition in the directed ER graph}
\label{sec:ER}

The concept of (undirected) random graphs was introduced 
by Erd\H{o}s and R\'enyi
 \cite{ER} in the 1950s in a simple model consisting of $N$ nodes 
and connecting
 every pair of nodes independently with the same probability $p$.
 Even though real networks differ from this
 simple model in many aspects, the ER graph
remains still of great interest,
since such a graph can serve both
as a test bed for checking all sorts of new ideas concerning
complex networks in general, and
as a prototype of random graphs to which all other random graphs can be
compared.

Perhaps the most conspicuous early
result on the ER graphs was related to the percolation transition
taking place at $p=1/N$.
The appearance of a {\em giant component} in a
network, which is also referred to as the {\em percolating component},
results in a dramatic change in the overall topological features
of the graph and has been in the centre of interest for
other networks as well. In a more general framework, one can also address 
the question of  $k$-clique percolation in the ER graph. 
Simple theoretical arguments as well as numerical simulations 
\cite{DerenyiPRL} show that  the critical linking probability of  
 $k$-clique percolation is
 $\pc^{\rm undir}=[(k-1)N]^{-1/(k-1)}$. In this section we carry out a similar
 analysis concerning the percolation transition of directed $k$-cliques
 in the directed ER graph.

\subsection{Derivation of the critical point}
The directed equivalent of the ER graph consists of $N$ nodes providing
 $N(N-1)$ possible ``places'' for the directed links, 
and these are filled independently
 with uniform probability $p$, producing on average $M\simeq N(N-1)p$ edges.
(Note that in the original undirected ER graph there are only $N(N-1)/2$
 possibilities to introduce an edge, therefore, at linking probability
$p$, there are  only $M\simeq N(N-1)p/2$ connections). The critical linking 
probability
 $\pc$ is decreasing with increasing $N$, and converges to zero as 
$N\rightarrow\infty$. We restrict ourself to the large $N$ limit, and
 evaluate $\pc$ to leading order only. Let us suppose that we approach
 the critical point from below: the directed $k$-cliques do not assemble
 yet into a giant module, we can find only small, isolated modules,
 and the system is dispersed. In terms of our $k$-clique template rolling
 picture this means that when trying to explore the directed percolation
 clusters by rolling such a template on them, we must stop the rolling 
 after a few steps as we run out of unexplored adjacent  directed $k$-cliques. 

 One can estimate $\pc$ from the condition that at the critical point 
the average number of yet unexplored directed 
 $k$-cliques adjacent to the $k$-clique we have just reached becomes equal 
to one. 
(This makes it possible to roll our template on and on for a long time). 
Since we are going
 to evaluate $\pc$ to leading order only, we can neglect the
 possibility to roll our $k$-clique template using double edges
 between the same nodes: When reaching a directed $k$-clique, 
 the minimal number of further edges that must be present to 
 enable the continuation of the template rolling is $k-1$. The probability
 of such a case is therefore proportional to $p^{k-1}$. Even though
 it is not forbidden in the first place to continue using double edges as
 well, each double edge in the new directed $k$-clique we are going 
 to roll onto multiplies the probability by $p$. In other words, the
 probability to roll further to a $k$-clique containing one double edge is
 smaller by a factor of $p$, the probability to roll further to a $k$-clique
 containing two double edges is smaller by a factor of $p^2$, {\it etc}. 

During the branching process exploring a directed $k$-clique percolation
 cluster, at the point when we are about to roll our template further on,
 we can choose the next node for relocation in $k-1$ different ways, which
 can then be relocated to approximately $N$ places.
If there were no restrictions
 for the directioning of the links inside a directed $k$-clique, then the 
 $k-1$ new links connecting the new node to this $k-1$ shared nodes could be
 directed in $2^{k-1}$ ways. However, the new directed $k$-clique has to
 fulfil the three condition detailed in Section 
\ref{sect:directed_k_cliques}. as well, therefore the actual number of 
allowed configurations is much smaller. The rank of the new node in 
 the new directed $k$-clique
 can be chosen in $k$ ways: the $k-1$ nodes shared with the previous $k$-clique
 are already ordered, and we can ``insert'' the new node to any place in this
 hierarchy. By fixing the order of the new node we fix
 the direction of the new links as well, therefore we can allow only $k$ 
different configuration for the directionality of these links.
By combining these factors together, the condition for reaching
 the critical point of the percolation transition can be written as
\bea
\pc^{k-1}N(k-1)k=1,
\eea
from which we gain
\bea
\pc^{\rm theor}=\left[Nk(k-1)\right]^{-1/(k-1)}=\pc^{\rm undir}/k^{k-1}
\label{eq:pc_theor}
\eea
for the theoretical prediction of the critical edge probability. Note that
 in the limiting case of $k=2$ (the directed edge percolation), the 
$\pc^{\rm theor}=\pc^{\rm undir}/2$ relation holds, which is consistent with the
2:1 ratio for the number of  links in the directed-- and undirected ER graph
 respectively.

\subsection{Numerical simulations}

\begin{figure}[t!]
\centerline{\includegraphics[angle=-90,width=1.1\columnwidth]{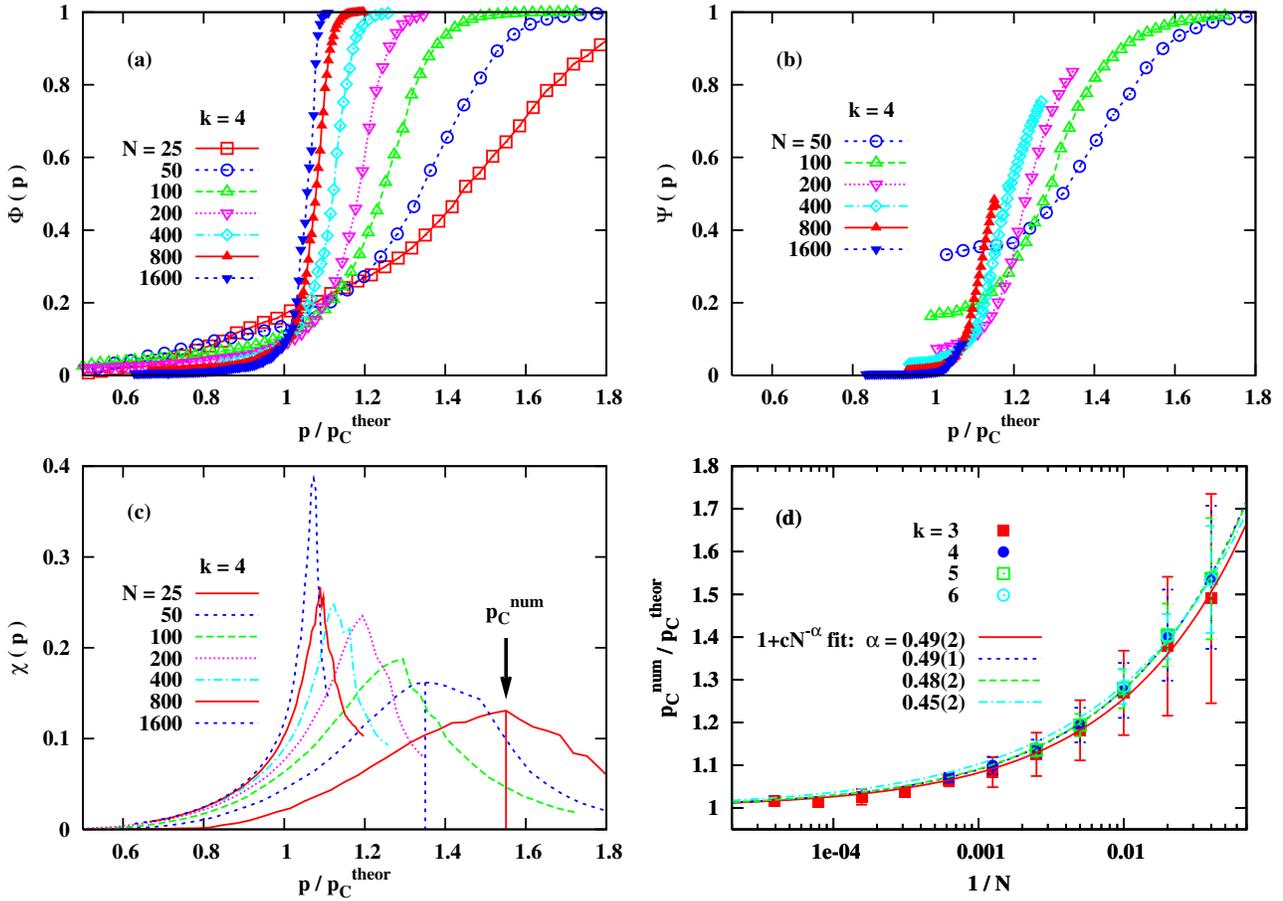}}
\caption[]{
Numerical results for directed $k$-clique percolation in ER-graphs. 
In each sub-figure, points show 
an average over $4$ to $100$ simulations depending on system size.
{\bf a)} The order parameter $\Phi$ (the number of nodes in the
 largest percolation cluster divided by $N$) as a function of
 $p/\pc^{\rm theor}$, where $\pc^{\rm theor}$ was obtained from Eq.\
(\ref{eq:pc_theor}). {\bf b)} The order parameter $\Psi$ (the number of
 directed $k$-cliques in the largest percolation cluster divided
 by the total number of directed $k$-cliques) as a function of 
 $p/\pc^{\rm theor}$. {\bf c)} The numerically determined value for the
 critical linking probability, $\pc^{\rm num}$, defined as
the average location of the maximum of $\chi(p)$, playing the role of the 
normalised percolation susceptibility (see Eq.\ \ref{eq:chi}).
{\bf d)} Verification of the theoretical prediction for the critical point. 
 The $\pc^{\rm num}/\pc^{\rm theor}$ ratio converges to one for large $N$.
}
\label{fig:ER}
\end{figure}

There are two plausible choices to measure the size of the largest
directed $k$-clique percolation cluster. The most natural one, 
which we denote by $N^*$, is the number of nodes belonging to this cluster. 
We can also
define an {\em order parameter} associated with this choice as the
relative size of this cluster:
\bea
\Phi=N^*/N.
\eea
The other choice is the number ${\cal N}^*$ of directed $k$-cliques
of the largest directed $k$-clique percolation cluster.
The associated order parameter is again the relative size of this
cluster:
\bea
\Psi={\cal N}^*/{\cal N},
\eea
where ${\cal N}$ denotes the total number of directed $k$-cliques in the 
graph. In Fig.\ref{fig:ER}a-b we display $\Phi$ and
 $\Psi$ as functions of $p/p_c^{\rm theor}$, where
the directed $k$-clique size is $k=4$, and
 the system size varies between $N=50$ and $N=1600$. The order parameter $\Phi$ 
converges to a step function for increasing system sizes, whereas $\Psi$ 
converges to a limit function (which is 0 for $p/\pc(k)<1$ and grows
continuously to 1 above $p/\pc(k)=1$). We have 
 evaluated the transition point numerically as well,
 by computing the second moment of the distribution of
$\mathcal{N}_i$ values, excluding the largest one,
$\mathcal{N}_1=\mathcal{N}^*$:
\bea
\chi = \sum_{i>1} \
\big(\,\mathcal{N}_i\,/\, \mathcal{N}\,\big)^2 \, .
\label{eq:chi}
\eea
\noindent
Note that this quantity is analogous to
the percolation susceptibility.
Both below and above the transition point
the $\mathcal{N}_i$ ($i>1$) values follow
an exponential distribution,
and only at $\pc$ do they have a power-law distribution.
Thus, $\chi$ is maximal at the numerical transition point, $\pc^{\rm num}$. 
In Fig.\ref{fig:ER}c we show $\chi$ calculated for the curves 
shown in Fig.\ref{fig:ER}b, as the function of $p/\pc^{\rm theor}$.
In order to check the theoretical prediction for the critical point
 obtained in (\ref{eq:pc_theor}) we have carried out a 
finite-size scaling analysis of the numerical results.
In Fig.\,\ref{fig:ER}d we show the ratio $\pc^{\rm num}/\pc^{\rm theor}$ as a
 function of $1/N$. Indeed, for large systems, the above ratio converges
 to one roughly as $1+cN^{-1/2}$.

\section{Results for real-world graphs}
\label{s_real}

In this section we study the directed modular structure of four real-world
 networks ranging from a word association graph through Google's web-pages
 to email and transcription regulatory networks.
When applied to real networks, the CPMd method has two parameters:
the $k$-clique size $k$, and (if the network is weighted) a 
weight threshold $w^*$ (links weaker than $w^*$ are ignored). 
Changing the threshold is like changing the resolution (as in a
microscope) with which the modular structure is investigated: by
increasing $w^*$ the modules start to shrink and fall apart. A very
similar effect can be observed by changing the value of $k$ as well:
increasing $k$ makes the modules smaller and more isolated from
 each other, but
at the same time, each module  becomes more cohesive.
When we are interested in the modular structure around a
particular node, it is advisable to scan through some ranges of
$k$ and $w^*$, and monitor how the obtained modules change. 
Meanwhile, when analysing the modular structure of the entire
network, the criterion used to  fix these parameters is based on finding a
 modular structure as highly structured as possible 
\cite{palla05nature}. This can be
 achieved by tuning the parameters just below the critical point 
 of the percolation transition. In this way we ensure that we
find as many modules as possible, without the negative effect of
having a giant module that would smear out the details of the
modular structure by merging (and making invisible) many smaller
modules. The technical details of the extraction of the directed
 $k$-clique modules are described in Appendix B.


\subsection{Word association graph}
\label{subsec:word}
\begin{figure}[t!]
\centerline{\includegraphics[angle=0,width=0.68\columnwidth]{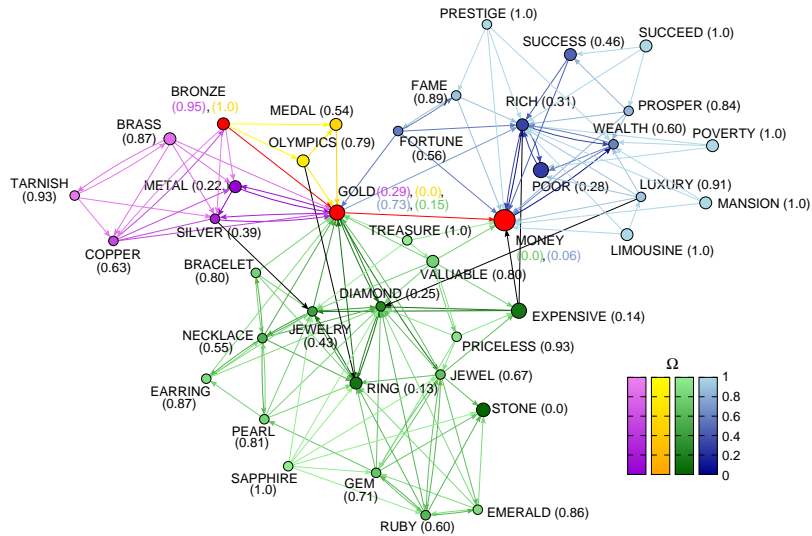}}
\caption[]{
The directed modules of the word ``GOLD'' at $k=4$, $w^*=0.023$ in
 the word association network \cite{word-assoc}.
The modules are colour coded and the overlaps between the modules
 are displayed in red. The size of each node is proportional to the
 number of modules it participates in (some of them are not shown in this
 figure). Beside the
 name of the nodes we display their
$\outnessw_{i,{\rm out}}^{\alpha}=
w_{i,out}^{\alpha}/(w_{i,in}^{\alpha}+w_{i,out}^{\alpha})$ values as well. 
Nodes with high $\outnessw$ (\eg ``SAPPHIRE'') usually correspond to special,
 rarely used words, whereas nodes with low relative out-degree 
(\eg ``MONEY'') are very  common.}
\label{fig:gold}
\end{figure}
\begin{figure}[t!]
\centerline{\includegraphics[angle=0,width=0.68\columnwidth]{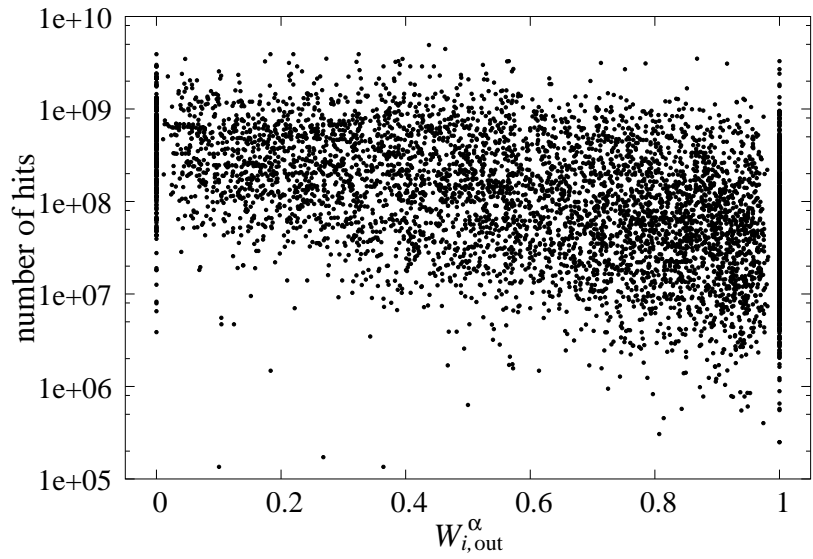}}
\caption[]{
The number of hits obtained from Google for module members as a function 
of their relative out-strength in the word association network. 
The number of hits is decreasing with increasing
 $\outnessw_{i,{\rm out}}^{\alpha}$, therefore, frequently used words are
 likely to obtain a low relative out-strength.
}
\label{fig:num_hits_rank}
\end{figure}

We  examined the directed network obtained from the
South Florida Free Association norms list (containing
 10617 nodes and 63788 links), where the weight of a 
directed link from one word to another indicates
the frequency that the people in the survey associated the end point of
the link with its start point \cite{word-assoc}. 
For illustration in Fig.\ref{fig:gold}. 
we show the (colour coded) modules 
 of the word ``GOLD'' obtained at $k=4$ and $w^*=0.023$, with the overlaps 
emphasised in red. According to its different meanings, this word participates
in four, strongly internally connected modules. Beside
 the node labels we display the relative out-strength of the nodes 
 in the modules using
 (\ref{eq:outnessw}). Apparently, nodes with a special/particular meaning 
 (\eg ``SAPHIRE'') tend to get high relative out-strength
whereas commonly used words with general meaning (\eg ``MONEY'') have
 low relative out-strength.
 Thus, it seems that the overall directionality of the modules
 is from special words towards more  general words. To make this
 observation more quantitative, we measured the number of hits obtained
 for the different words appearing in the network using the search
 engines of Google. In Fig.\ref{fig:num_hits_rank}.
 we show the scatter plot of the number of hits as a function of 
 the relative out-strength of the members of the modules obtained at the
optimal $k=4$, $w^*=0.016$ parameters. The decreasing tendency of 
the number of hits with 
increasing $\outnessw_{i,{\rm out}}^{\alpha}$ signals that words 
with higher relative out-strength
(\ie, having mainly out-neighbours) are usually
 less frequently used than words with lower relative out-strength
(\ie, having mainly in-neighbours).

\subsection{Google's web-pages}
\label{subsec:google}

\begin{figure}[t!]
\centerline{\includegraphics[angle=0,width=0.67\columnwidth]{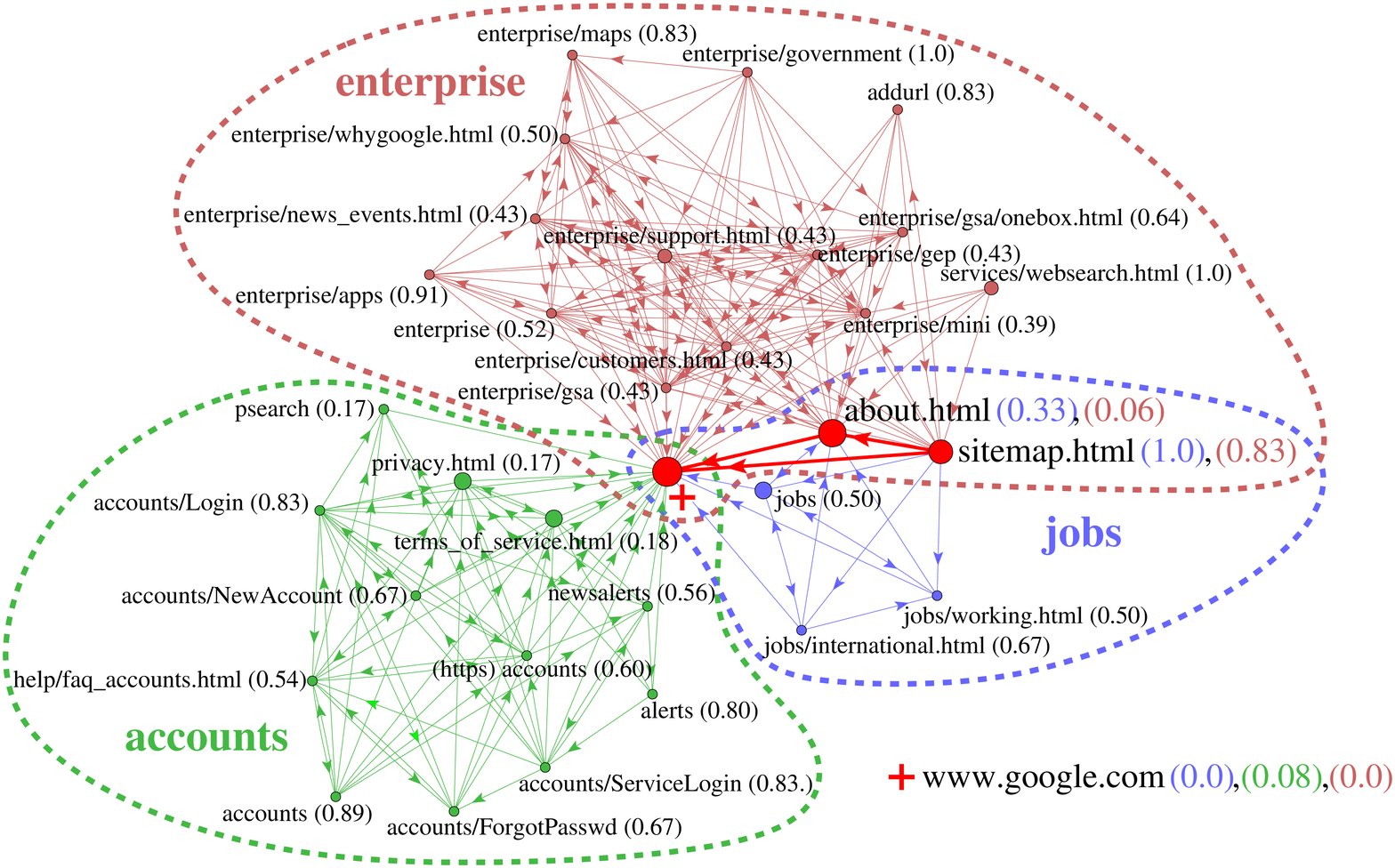}}
\caption[]{
Three of the overlapping directed modules identified by CPMd 
in the directed net of Google's static pages at $k=6$.
These modules overlap with several further ones not shown in the figure; 
the size of each node is proportional to the number of its modules.
The nodes and links of the three modules are coloured brown, green and blue, 
while their overlaps, \ie, nodes contained by more than one of these 
three modules, are red. The node marked with a {\bf +} sign at centre 
is the starting page, http://www.google.com, and the names of the
other nodes are their URLs without this prefix. The $\outness$ 
 values of the module members are marked beside the node labels. 
Observe that each module 
contains a number of nodes with many incoming links (a ``core''), 
some of which are in the overlaps. See text for further details and 
Fig.\,\ref{fig:hubCompare} for a detailed analysis of hubs and overlaps.
}
\label{fig:Google}
\end{figure}

In addition to being a prominent means of information retrieval,
Google provides its own documents as well: usage notes, feature 
and product descriptions, \etc. The map of hyper-links among Google's 
own web-pages offers a unique insight into how one of the major search 
portals arranges online content and thereby helps and guides our browsing.
Excluding dynamic content and catalogues, we downloaded 
with our robot \cite{robot} $15,763$ web-pages (nodes) and $171,206$
directed links among them. For the current analysis,
we excluded international pages and nodes farther than $3$ steps 
from the start node, http://www.google.com,
and obtained a graph with $946$ nodes and $1,817$ links.
Fig.\,\ref{fig:Google}. shows three of the many overlapping directed modules 
identified by the CPMd in this network at $k=6$.
 Apparently each of the identified overlapping modules 
in Fig.\,\ref{fig:Google} is a group of internally densely 
connected nodes organised around a well-defined topic
(jobs, accounts and enterprise solutions).

An interesting feature of Google's directed modules is that
they {\it share their in-hubs, but not their out-hubs.} (By ``in-hub'' we
 mean nodes with outstanding in-degree, whereas ``out-hub'' stands for
 nodes with outstanding out-degree).
This structure enhances browsing efficiency.
Having visited a particular, ``outlying'' page of a module,
one can quickly return
to a node in the core of the same module.
Then, due to the strong overlaps among the cores,
one can quickly jump over to a new topic, \ie,
the web-pages of another module.
In summary, 
our ability to browse efficiently and hierarchically
Google's web-pages is enhanced by the facts that
modules overlap via their in-hubs.

\subsection{Email network}
\label{subsec:email}

\begin{figure}[t!]
\centerline{\includegraphics[angle=0,width=0.98\columnwidth]{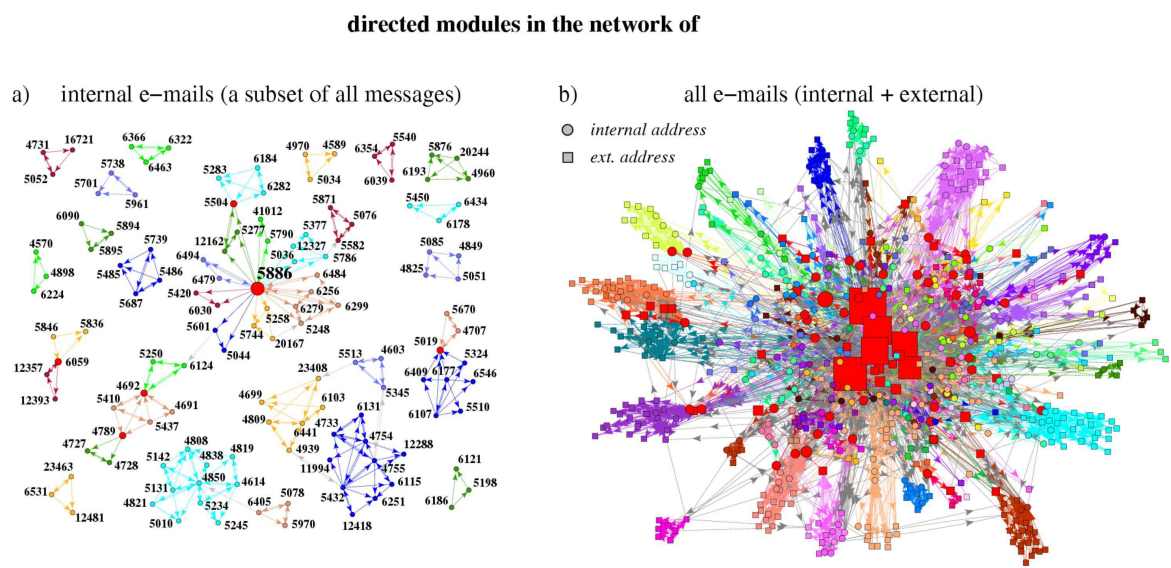}}
\caption[]{
All directed modules in a network of student emails
at the University of Kiel
during a period of $112$ days (data from Ref.\,\cite{email}).
On the left {\bf (a)} only the graph of internal emails 
(between students of the university) is analysed,
while on the right {\bf (b)} 
internal and external messages are both included.
Circles and boxes show internal and external email addresses,
respectively, and the size of a node is proportional to the number
of its modules. The largest nodes, \ie, those with the highest 
membership number, have significantly more outgoing than incoming
links, meaning that in this email network modules 
share their out-hubs.
See also Fig.\ref{fig:hubCompare}.
The optimal $k$-clique size parameter values are $k=3$ (a) and $k=4$ (b).
}
\label{fig:email}
\end{figure}

A very common type of directed social networks is the one defined by
messages and information flow (directed links) among individuals (nodes).
To ``measure'' such a social network, Ebel 
and Bornholdt \cite{email} processed the directed network defined
by the emails of students at the University of Kiel 
during a period of $112$ days.
We analysed both the entire data set and its subset containing only 
emails between internal addresses (students). The full network contains 
$57,158$ nodes and $103,701$ links, while the $1,267$ internal addresses 
(nodes) are connected by $1,659$ links. Fig.\,\ref{fig:email} 
shows the directed modules in these two
networks. Observe that even among
the relatively small number of internal emails modules, overlaps 
do appear, \eg, node 5886 at the centre.
In the full e-mail data set, external addresses have both the highest
degrees (number of connections) and the largest numbers of modules they
participate in. 
In contrast to \eg the Google's web-pages, nodes with the largest out-degrees 
participate in a high number of modules.

\subsection{The transcriptional regulatory network in yeast}
\label{subsec:yeastTR}

In a cell the transcription of a gene is influenced (regulated) by one or
more proteins called transcription factors. This regulatory relationship 
is most often represented as a directed link pointing from the regulating 
protein (source node) to the protein of the regulated gene (target node).
Recent experimental and computational techniques \cite{Harbison,MacIsaac}
have enabled the genome-wide mapping of transcription regulatory
relationships in the yeast, {\it S. cerevisiae}.

\begin{figure}[t!]
\centerline{\includegraphics[angle=0,width=0.77\columnwidth]{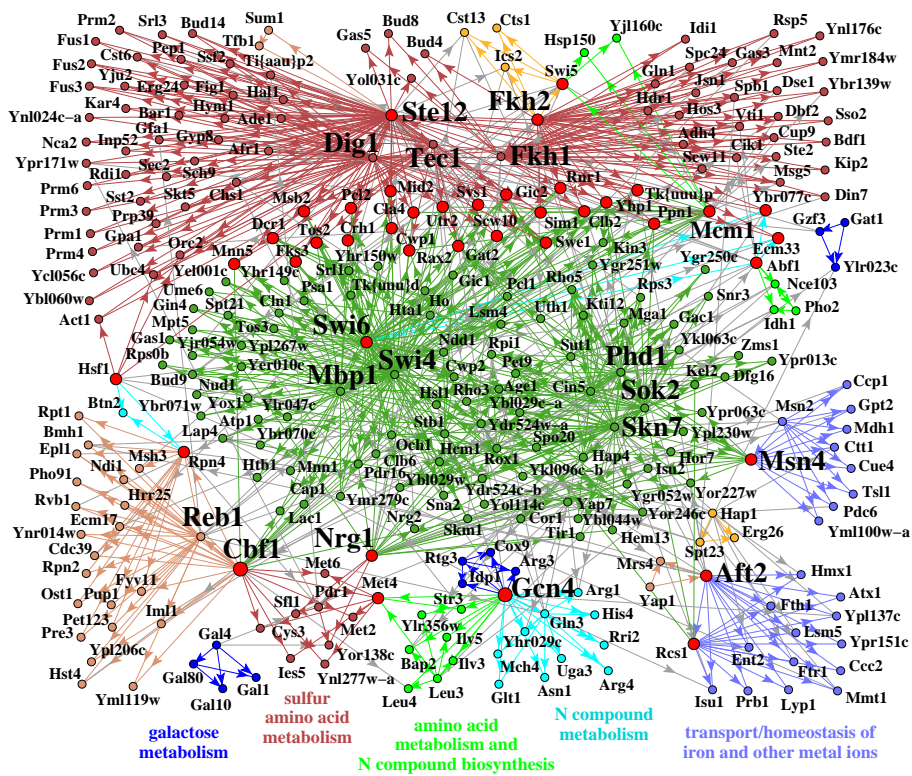}}
\caption[]{
The directed modules of the web of transcription regulatory
interactions in baker's yeast ($k=3$).
Each node shows one gene (and its protein) and a directed link stands
for a transcription regulatory interaction between a protein and the
target gene. Modules (communities) are coloured and overlaps
are red. The overlapping nodes are mostly out-hubs.
Group functions have been identified by GO TermFinder \cite{GOtermfinder}.
}
\label{fig:yeastTR}
\end{figure}

In Fig.\ref{fig:yeastTR}. we display the obtained directed modules 
for $k=3$. As an example, for some of the modules the most significant common
functions of their participating proteins have been identified 
from the Gene Ontology protein function annotation database \cite{GO}
with the search tool GO TermFinder \cite{GOtermfinder}. The list of 
regulatory interactions  was obtained from Ref.\,\cite{MacIsaac}. 
Most  protein  modules in Fig.\ref{fig:yeastTR} are arranged around 
a small number of large out-hubs, the major transcription factors (TFs), 
each of which regulates a large portion of all target genes in the module.
Overlaps between the modules occur either through the TFs, \eg, 
via the nodes Met4 and Gcn4 in the bottom left part of the figure, 
or via large groups of regulated (target) genes, see, \eg, the red nodes 
at the ``interface'' between the yellow and brown modules in the upper 
part of the figure. Hence, from the point of view of directed modules, 
the transcription regulation network is organised in a similar way to 
the e-mail network, and an opposite way to Google's web-pages (and the
 word association network).

\subsection{Comparison between CPMd and CPM}
\label{subsection:compare}

For each studied network, by ignoring the directionality of the links, 
we located the CPM communities as well. In case of the word association
 network, where links are weighted as well, the weight of the 
undirected counterpart of a double link was defined as the sum of the
 corresponding two weights. Due to this difference in the weights as well
 as in the definition of modules, the optimal weight threshold was
 slightly different in the CPM approach. 

Surprisingly, in spite of the restrictions of the CPMd compared to CPM, 
(and in case of the word association network, the difference in link weights),
 about 70\% of the modules were the same in the two approaches for
 the word association network and Google's web pages, whereas this ratio
 turned out to be even higher (around 90\%) for the email network
 and the transcription regulatory graph. Furthermore, for the rest of the 
directed modules one could find a relatively similar undirected module in
 most of the cases. This shows that the original CPM approach to the
identification of modules
 is quite robust, our restrictions introduced in the CPMd leave the 
 majority of the undirected modules intact.

\subsection{Classification of real-world networks:
modules are connected by in-hubs or out-hubs}
\label{ss_hubs}

An important aspect of network motifs (overrepresented
small sub-graphs with a given structure) is that complex networks can
 be classified based on their motif significance profile (a pattern 
of motif usage) \cite{milo04}. In a somewhat similar approach, here
 we classify the four investigated real-world webs into two major 
groups based on the overlaps of their directed modules.

\begin{figure}[t!]
\centerline{\includegraphics[angle=-90,width=0.67\columnwidth]{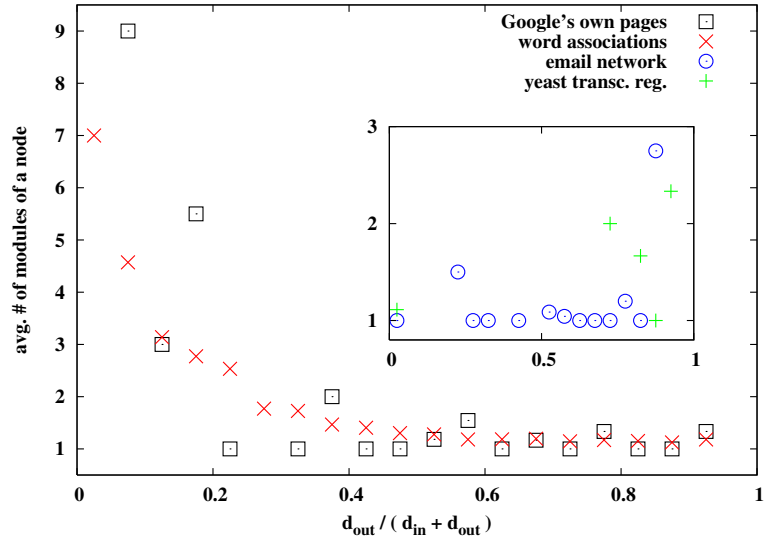}}
\caption[]{
The average membership of a node vs. its 
$d_{\rm out}/(d_{\rm in}+d_{\rm out})$ ratio. 
This function is a growing (decreasing) one, if the modules are
more likely to overlap via in-hubs (out-hubs).
}
\label{fig:hubCompare}
\end{figure}

Interestingly, the way that the out-hubs  and in-hubs
of the network are arranged within its directed modules is 
different among the various types of networks. To directly 
compare the studied networks from this aspect, in 
Fig.\ref{fig:hubCompare}. we show the average number
 of modules of the nodes as a  function 
of their relative out-degree $\outness_{i,{\rm out}}\equiv d_{i, {\rm out}}/(d_{i, {\rm in}}+
d_{i,{\rm out}})$ ratio. Apparently,
the modules in the word association network and Google's web-pages are
connected by in-hubs: nodes contained by a large number of modules have
 a small $D_{i,{\rm out}}$. In contrast, in the
 email network and the transcription regulatory graph of yeast the overlaps
are more likely to contain out-hubs than in-hubs.

The plausible reason for the observed difference between the investigated
 networks is that overlaps contain hubs with increased likelihood in
 the first place, and the two kinds of hubs occur in the networks 
with different probabilities. 
In the word-association network and Google's web 
pages in-hubs are more frequent: the number of words
 we associate to a cue word and the number of hyper-links that appear on
 a web page is more or less constant, however a word with a general meaning
 or an important (general) web page can appear as the target for many 
links. In contrast, we are more likely to find out-hubs than in-hubs in 
 the email network and the transcription regulatory graph. The time spent
on sending an email does not depend on the number of recipients, whereas 
reading a large number of incoming emails can take a lot of time, therefore
 being an in-hub in the email network is disadvantageous and in-hubs are
 rare. Similarly, in case of the transcription regulatory graph the number of
 transcription factors that can regulate a given protein is more or 
 less constant, whereas a single transcription factor can regulate
 many other proteins in parallel, therefore, out-hubs are much more
 frequent than in-hubs.

\section{Summary and conclusions}

We examined the directionality of network modules. To compare and order
 the nodes in a module, we  introduced the relative out-degree, 
measuring the relative weight of the out-links of a member to other 
nodes in the module. We developed a specific module finding algorithm for
 directed networks as well, based on the $k$-clique percolation approach.
Even though the CPM can be extended to any kind of directed $k$-cliques 
(containing an arbitrary set of directed links), here we concentrated 
on the most plausible choice which
allows a straightforward theoretical and numerical analysis.
Following a simple branching procedure, we have derived the critical point 
of the directed $k$-clique percolation in the ER graph in the large $N$ 
 limit. The theoretical prediction was justified by numerical simulations. 
We have also studied the directed modular structure of real-world networks
including a word association graph, Google's web pages, an e-mail network
 and the transcription regulatory network of yeast.  The obtained modules
 were validated by additional information (annotations) for the members.
The nodes contained in the overlaps between the modules enabled
us to classify the examined networks in two major groups: 
the modules in the word association graph and Google's web pages are
likely to be connected by in-hubs, whereas the overlaps in the e-mail network
 and the transcription regulatory network are more likely to contain
out-hubs.

\ack

The authors thank the partial support of the Hungarian National Science
Fund (OTKA T034995, K068669, PD048422) and the National Research
 and Technological Office (NKTH, CellCom RET).

\section*{Appendix A}
In this appendix we show that for $k$-cliques with no double links, the 
following three statements are equivalent:
\begin{itemize}
\item[(i)] Any directed link in the $k$-clique points from a node with a
  higher order (larger restricted out-degree) to a node with a 
lower order. 
\item[(ii)] The $k$-clique contains no directed loops.
\item[(iii)] The restricted out-degree of each node in the $k$-clique 
is different.
\end{itemize}
(The restricted out-degree of a node is equal to the number of its 
out-neighbours in the $k$-clique).\\

\noindent (ii)$\rightarrow$(iii) : {\it If loops are absent, 
then all the members have different restricted out-degrees.}
If there are no loops, then there must be a node in the $k$-clique having all
 in-neighbours amongst the other members, since otherwise we could 
hop from node
 to node following a directed link inside the $k$-clique forever, (which would
 mean that it does contain at least one loop). If we reversed the 
direction of all links inside the $k$-clique we would not induce any loops, and
 therefore, this "reversed" configuration would have a member with only 
incoming links from the others as well. From this it follows that there must be
 also a node in the $k$-clique with only out links towards the other nodes.  
By removing this node we obtain a $(k-1)$-clique in which directed loops are 
absent. Similarly to the previous case, this $(k-1)$-clique must have a node 
with only out-neighbours amongst the other members of the $(k-1)$-clique. By 
removing this node as well, we arrive at a $(k-2)$-clique containing no loops.
 And so on, by subsequently removing the node with only out-neighbours at each 
step we iterate over all nodes, and obviously the restricted out-degree of 
the removed node is decreased by one at each step, hence all nodes have 
different number of out-links inside the $k$-clique.\\ 

\noindent (ii)$\rightarrow$(i) :  {\it
If loops are absent, then the links point from higher
 restricted out-degrees values towards lower ones.}\\ 
The above process showing  (ii)$\rightarrow$(iii) also reveals 
that the links inside a $k$-clique with no loops are always
 pointing from a node with a higher restricted out-degree towards a node 
with less out-links inside the $k$-clique.\\

\noindent (iii)$\rightarrow$(ii) : {\it
If all nodes have different number of out-neighbours inside 
the $k$-clique, then directed loops are absent.}\\
The possible number of out-neighbours a node can have inside a $k$-clique falls
 in a range between 0 and $k-1$, therefore, if all nodes have different number
 of out-neighbours, then all of these possible values must actually appear in 
the $k$-clique. Since double links are absent, the node with $k-1$ 
out-links cannot have any incoming 
links from the other members, therefore, it is surely not part of any directed
 loops inside the $k$-clique. The node 
with $k-2$ out links has only a single incoming link, starting at the node 
with only out-links. Therefore, this node cannot be part of any directed loops
 either. Similarly, the node with $k-3$ out links has two incoming links, 
both starting at nodes that have been already shown to be excluded from any 
directed
 loops (the nodes with $k-1$ and $k-2$ out-neighbours, respectively). Thus, 
the node with $k-3$ out neighbours amongst the other members "inherits" this
 property (to be excluded from directed loops inside the $k$-clique) as well.
 And so on, by subsequently scanning the nodes in decreasing order of their 
restricted out-degrees, at each step all the 
incoming links to the node under investigation come from previously examined
 members that were shown to be excluded from loops, therefore, the 
investigated node cannot be part of any loops either. \\

\noindent(i)$\rightarrow$(iii) : {\it If  each 
directed link points from a node
 with a higher restricted out-degree to a node with a lower one, then
 the restricted out-degree of each node in the $k$-clique is different.}\\
This statement is almost trivial, since if any pair of nodes had 
the same restricted out-degree, then the link
 connecting them would point in the direction of constant restricted 
out-degree.\\

For $k$-cliques with double links none of the three statement can hold. The
 presence of loops is trivial: a double link is already equivalent of
 a closed directed path. Furthermore, both constituents of a double link
 cannot point in the direction of decreasing order simultaneously. 
Therefore, we only have to prove that (iii) cannot be true either, \ie
for $k$-cliques with double links their
 members cannot have all different numbers of out-neighbours amongst the other
 nodes in the $k$-clique. The total number of links, $m$, inside a $k$-clique
 can be written as
\bea
m=\sum_{q=0}^{k-1}qn_q,
\label{eq:biz}
\eea  
where $q$ runs over the possible number of out-neighbours, and $n_q$ is the 
number of members with the given restricted out-degree.
 When all the members have different number of out-links, $n_q=1$ for all 
possible $q$ values, and thus, $m=k(k-1)/2$, which is exactly the number of
 links in a $k$-clique with no double links. However, in presence of double 
links $m>k(k-1)/2$, therefore, at least one of the $n_q$ values in 
(\ref{eq:biz}) must be larger than one, meaning that there are nodes in 
the $k$-clique with equal restricted out-degrees.  \\

\section*{Appendix B}

In this section we briefly describe our algorithm for extracting 
the CPMd modules in networks.
Since any subgraph of a directed $k$-clique is a directed $k$-clique as 
well, (with a smaller $k$ value), an efficient way to extract the
 directed $k$-clique modules of a network is to first find all
 {\it directed cliques} first: A directed clique is a maximal directed
 $k$-clique, \ie it is not part of an even larger directed $k$-clique.
 A CPMd module of a given $k$ is equivalent of the union of 
 directed cliques of size larger or equal to $k$, which can be
 reached from each other through overlaps of size larger or equal
 to $k-1$. 

\begin{figure}[h!]
{
\centerline{\includegraphics[width=0.68\textwidth]{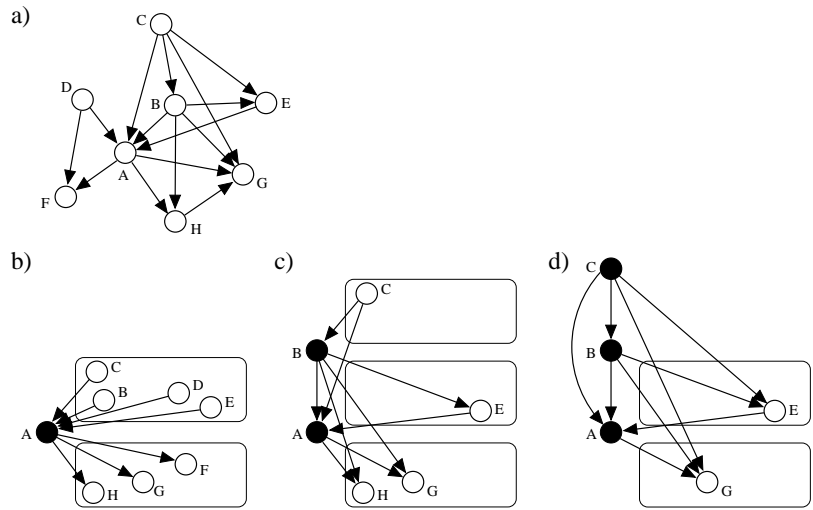}}}
\caption{
Illustration of the directed clique search. {\bf a)} The neighbourhood
 of node $A$ in a hypothetical directed network. {\bf b)} 
The initial state of the
directed clique extraction algorithm: the in-neighbours of $A$ are above 
 $A$, whereas its
 out-neighbours are below it. {\bf c)} 
Node $B$ is picked from the in-neighbours
 and is placed above $A$. Nodes $D$ and $F$ are not neighbours of $B$,
 therefore they are removed from the containers. Furthermore, a new container
 is introduced holding node $E$, which is in-between $B$ and $A$ in the
 hierarchy. {\bf d)} Node $C$ is picked from the top container and is placed
 above $B$, node $H$ is removed from the bottom container as it is not
 linked to $C$.
}
\label{fig:dir_cl_search_expl}
\end{figure}

We extracted the directed cliques using the following iteration
\begin{enumerate}
\item find all directed cliques of a given node,
\item remove the node and its links from the network.
\end{enumerate}
To find the directed cliques of a given node, $A$, we use a back-tracing 
algorithm based on the hierarchical properties of the directed cliques. 
At the initial step we construct two containers, one for the in-neighbours
and one for the out-neighbours of $A$. The hierarchy of the system at
 this point is illustrated in Fig.\ref{fig:dir_cl_search_expl}b: 
the in-neighbours are at the top,
 the out-neighbours are at the bottom, and the node $A$ itself is in-between
 them. Next we take a node from the in-neighbours (or the out-neighbours),
 this node and $A$ form a directed 2-clique. We place the node above
 (or below) $A$, and filter the remaining nodes in the containers so that
 for both nodes in the newly formed 2-clique it is true that
\begin{itemize}
\item the members in the containers above the node in the hierarchy are
 all in-neighbours of node $A$,
\item the members in the containers below the node in the hierarchy are
 all out-neighbours of node $A$. 
\end{itemize}
If necessary, we may introduce a new container as well, \eg, in 
Fig.\ref{fig:dir_cl_search_expl}c, by
 picking node $B$ from the top container, the node $E$ which is an 
out-neighbour of $B$ and an in-neighbour of $A$ is placed in a container
 in-between $B$ and $A$ in the hierarchy. This way when picking the next node
 from any of the containers, its rank in the hierarchy inside the forming 
 directed clique coincides with the rank of its container with respect to
 the already selected nodes. For example, when picking node $C$ in the example
 shown in Fig.\ref{fig:dir_cl_search_expl}, it is placed above node $B$. 
 By recursively picking new nodes
 from the containers, filtering the containers and introducing new containers
 we build up a directed clique. (The extraction of the clique ends when
 all containers become empty). 

Our algorithm scales similarly to the original CPM
 (see the Supplementary Information of 
Ref.\cite{palla05nature}).
Since the determination of the full set of cliques of a graph is
 widely believed to be a non-polynomial problem, the extraction
 of the directed cliques is non-polynomial as well. In spite of this,
in real networks 
our algorithm proves to be quite efficient. 
Our experience shows that the required
 CPU time depends on the structure of the input data very strongly,
 therefore, in general no closed formula can be given even to estimate the
 system size dependence. As an illustration of the computational speed,
 however, we note that a complete analysis of the word-association network
 with over 70,000 links takes less than 5 minutes on a PC. 
By extracting the directed modules of this system at different link-weight
 thresholds, the time dependence of the algorithm could be
 fitted with $t=AM^{B\ln(M)}$ where $t$ denotes the time needed 
 by our program,
 $M$ stands for the number of edges, and $A$ and $B$ are fitting parameters.

\end{document}